\documentclass[11pt]{article}
\usepackage{geometry}   
\usepackage{dsfont}
\usepackage{amsthm}
\usepackage{amsmath}
\usepackage{amssymb}
\usepackage{esint}
\usepackage{graphicx}
\usepackage{mathrsfs}       
\usepackage{graphicx}
\usepackage{amssymb}
\usepackage{epstopdf}
\usepackage{textpos} 
\usepackage{color}
\usepackage[dvipsnames]{xcolor}
\usepackage{subfigure}
\usepackage[font={small,it}]{caption}
\usepackage[makeroom]{cancel}
\DeclareMathOperator{\blueb}{B}
\DeclareMathOperator{\redr}{R}

\title{Entanglement Enabled Intensity Interferometry \\ of Different Wavelengths of Light} 
\author{Jordan Cotler$^1$,  Frank Wilczek$^{2,3,4,5,6}$ and Victoria Borish$^7$}
\date{}

\begin{document}
\maketitle
\vskip-1cm
\begin{textblock*}{5cm}(12.5cm,-5.2cm)
  \fbox{\footnotesize MIT-CTP-4814}
\end{textblock*}
\begin{center}
\small\it 1. Stanford Physics Department, Stanford University, Stanford CA 94305 USA \\
\small\it 2. Center for Theoretical Physics, MIT, Cambridge MA 02139 USA \\
\small\it 3. T. D. Lee Institute, and Wilczek Quantum Center, Shanghai Jiao Tong University, \\ Shanghai 200240, China\\
\small\it 4. Department of Physics and Origins Project, Arizona State University, Tempe AZ 25287 USA \\
\small \it 5. Department of Physics, Stockholm University, AlbaNova, 10691 Stockholm, Sweden \\
\small \it 6. Nordita, KTH Royal Institute of Technology and Stockholm University, Roslagstullsbacken 23, 10691 Stockholm, Sweden \\
\small\it 7. Department of Applied Physics, Stanford University, Stanford CA 94305 USA
\vskip1cm
\end{center}

\begin{abstract}
We propose methods to perform intensity interferometry of photons having two different wavelengths.
Distinguishable particles typically cannot interfere with each other, but we overcome that obstacle by processing the particles via entanglement and projection so that they lead to the same final state at the detection apparatus.  Specifically, we discuss how quasi-phase-matched nonlinear crystals can be used to convert a quantum superposition of light of different wavelengths onto a common wavelength, while preserving the phase information essential for their meaningful interference.  We thereby gain access to a host of new observables, which can probe subtle frequency correlations and entanglement.  Further, we generalize the van Cittert-Zernike formula for the intensity interferometry of extended sources,
demonstrate how our proposal supports enhanced resolution of sources with different spectral character, and suggest potential applications.   

\end{abstract}
\vskip1cm
\noindent \textit{Note: This paper is a combination of:} \\ \\
Cotler, Jordan, and Frank Wilczek. ``Entanglement Enabled Intensity Interferometry." arXiv:1502.02477v1 (2015). \\ \\
Cotler, Jordan, Frank Wilczek, and Victoria Borish. ``Entanglement Enabled Intensity Interferometry of Different Wavelengths of Light.'' arXiv:1607.05719v1 (2016).
\newpage
\section{Introduction}
The Hanbury-Brown Twiss (HBT) effect is a staple of multi-particle interferometry, with broad applications in experimental physics \cite{HBT1, HBT2, HBT3, HBbook, Baym1, Csernai1}.
It is generally considered that indistinguishability of particles is a central requirement for multi-particle interference, but in fact it is possible to generalize HBT
to allow for the interference of distinguishable particles.  Interference between potentially distinguishable particles can be observed by a using detection apparatus that does not, at a fundamental quantum mechanical level, distinguish the particles: a procedure we call
``Entanglement Enabled Intensity Interferometry" (E$^2$I$^2$).

To enable interference between photons of different wavelength, it is not enough that the detector fail to read out the wavelength of an incoming photon.  Rather, the detector must reach the same final quantum state in response to photons of either wavelength.   To achieve that goal, one must exploit appropriate projections and/or
entanglement between the incoming particles and the detector.  In this paper, we explore several methods involving nonlinear crystals, coherent pumps, and filters to demonstrate intensity interferometry with photons of different wavelengths.
Our methods potentially have applications in several fields, notably including astronomy and fluorescence microscopy, in which one might profit from improved resolution among sources with different spectral characteristics.

Nonlinear optical properties of crystals figure centrally in our practical proposals. Frequency upconversion and downconversion are  forms of three-wave mixing in which the frequency of a photon is changed while its other quantum properties (including phase information) are preserved \cite{Raymer1, Kumar1, Kumar2}. In these processes, an input photon as well as a coherent state (referred to as the pump) interact in the nonlinear crystal in such a way that the frequency of the input photon is altered while a pump photon is created or annihilated, conserving energy.
Quantum frequency upconversion has been used to bring the wavelength of a photon into a region where detectors are very efficient \cite{Albota, Langrock}, while quantum frequency downconversion may be used to convert photons from a wavelength convenient for information processing to a wavelength that is convenient for transmission \cite{Ou, Takesue2, Curtz, Fejer2}.

The possibility to perform frequency conversion on sources of different wavelength, thus producing indistinguishable photons which subsequently interfere, has been demonstrated experimentally. This was achieved in the context of the Hong-Ou-Mandel dip, where two photons of different wavelength were both upconverted before arriving simultaneously at a beamsplitter \cite{Takesue1}.  The quality of the resulting interference quantifies the success of the upconversion and its ability to preserve phase information.  It is even possible to combine frequency conversion and beamsplitting into a single step \cite{Raymer2}.

In HBT, two detectors can receive photons from either of two distinct sources.  The inability of the detectors to determine which photons come from which source is a necessary condition for two-photon interference.  
If the sources emit distinguishable photons, it becomes possible to determine which photons were emitted from which source.  If our detectors are capable of making that determination, then interference becomes impossible.  In many applications of interferometry, one only has access to light from a particular source after it has been spatially mixed with light from other sources.  To address such applications we must modify the procedures of the previous paragraph.  A major goal of this paper is to spell out how to do that.   

In Section \ref{sec:reviewE2I2}, we explain E$^2$I$^2$ in the context of HBT and describe the mechanism that allows one
to perform intensity interferometry on sources of different wavelengths conceptually. Then in Section \ref{sec:Mechanism}, we detail three different methods, of increasing complexity and generality, for implementing this mechanism for E$^2$I$^2$.   Section \ref{sec:implementation} provides a detailed, concrete proposal for a proof-of-principle experiment demonstrating the first of the methods from Section \ref{sec:Mechanism}.  Applications are described mathematically in Section \ref{sec:resolution}, where we generalize the classic van Cittert-Zernike formulae, and specific examples are given in \mbox{Section \ref{sec:applications}}.  Additional protocols for E$^2$I$^2$ in more general settings are presented in the Appendix.

\section{E$^2$I$^2$ for HBT} \label{sec:reviewE2I2}

\begin{figure}[h]
\centering
\includegraphics[scale=0.6]{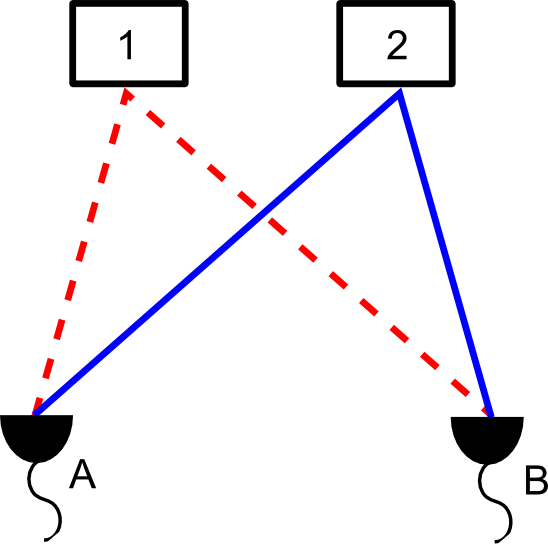} 
\caption[]{Geometry of the Hanbury-Brown Twiss intensity interferometer.  Two distinct processes contribute to correlated firing of the detectors.}
\label{fig:hbtGeometry} 
\end{figure}

In this section we explain E$^2$I$^2$ in the context of HBT, and establish our notation.  Consider the setup displayed in Fig. \ref{fig:hbtGeometry}.

\noindent We have two sources, labeled $1$ and $2$, which emit red and blue photons, respectively.  At the other end of the setup, we have two detectors $A$ and $B$.  The propagator (i.e., transition amplitude) from 1 to $A$ is denoted by $D_{1A}$, and the other propagators for source-detector combinations are defined similarly.  Temporarily neglecting the possible difference in arrival times for simplicity, suppose that each source emits a photon that propagates towards the detectors.  Then the probability of coincidence between the two detectors is
\begin{equation}
\label{noint1}
\Big| \, D_{1A}D_{2B} \, |\redr \! \blueb \rangle \, + \, D_{2A}D_{1B} \, |\blueb \! \redr \rangle \,\Big|^2 = |D_{1A} D_{2B}|^2 + |D_{2A} D_{1B}|^2
\end{equation}
where $|\redr \! \blueb \rangle $ is the state of the system when a red photon is absorbed by detector A and a blue photon is absorbed by detector B, with $|\blueb \! \redr \rangle$ defined similarly.  Eqn. (\ref{noint1}) does not exhibit interference between the $D_{1A}D_{2B}$ and $D_{2A}D_{1B}$ terms, which follows from $\langle \redr \! \blueb | \blueb \! \redr \rangle = 0$ since $|\redr \! \blueb \rangle $ and $|\blueb \! \redr \rangle$ are distinguishable states.

Now suppose that at each detector we implement a unitary transformation $U$ which maps
\begin{align}
|\blueb\rangle &\longrightarrow \cos(\theta) \, |\blueb\rangle + e^{i \phi} \, \sin(\theta)\, |\redr\rangle \\
|\redr\rangle &\longrightarrow e^{- i \phi} \sin(\theta) \, |\blueb\rangle + \cos(\theta)\, |\redr\rangle
\end{align}
and then filter out all states which are not $|\blueb\rangle$.  This corresponds to computing
$$|\blueb \! \blueb \rangle  \langle \blueb \! \blueb | \, (U \otimes U) \, \left( D_{1A}D_{2B} \, |\redr \! \blueb \rangle \, + \, D_{2A}D_{1B} \, |\blueb \! \redr \rangle \right)$$
and then renormalizing the state to obtain
$$D_{1A}D_{2B} \, |\blueb \! \blueb \rangle \, + \, D_{2A}D_{1B} \, |\blueb \! \blueb \rangle .$$
In this situation the probability of coincidence, given that the blue photons have successfully passed through the filter, becomes
\begin{align}
\label{int1}
\big| \, D_{1A}D_{2B} \, |\blueb \! \blueb \rangle \, + \, D_{2A}D_{1B} \, |\blueb \! \blueb \rangle \, \big|^2  =  | D_{1A} |^2 \, | D_{2B}|^2 \, + \, | D_{2A}|^2 \, |D_{1B} |^2 \, + \, 2 \, {\rm Re} \, D_{1A}D_{2B}D_{2A}^*D_{1B}^* \nonumber \\
\end{align}
which contains the interference term $2 \, {\rm Re} \, D_{1A}D_{2B}D_{2A}^*D_{1B}^*$.  From this toy model, it becomes clear that if we want to recover the interference term, our task is to design a detection apparatus which implements the unitary $U$ on incoming light and then applies the applies the appropriate filter.  In particular, we would like $U$ to allow for the possibility that each of two different wavelengths can be converted to a common wavelength without changing the other properties of the light.  In the next section, we provide a description of how to implement such a unitary.

\section{Mechanism for E$^2$I$^2$ with Different Wavelength Photons} \label{sec:Mechanism}
\subsection{Single Crystal Method}
Consider photons of wavelength $\lambda_1$ and $\lambda_2$ with energies $E_1$ and $E_2$, respectively.  Further, we suppose that $E_1 < E_2$ and define $\Delta E := E_2 - E_1$.  Photons with energy $\Delta E$ will be denoted by their wavelength $\lambda$. 

\begin{figure}[h]
\centering
\includegraphics[scale=1.5]{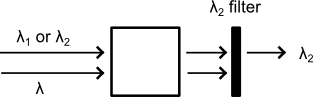} 
$$$$
\caption[]{A photon of wavelength $\lambda_1$ or $\lambda_2$ is transformed and projected onto a photon at wavelength $\lambda_2$. This can be achieved, for example, when the photon of wavelength $\lambda_1$ or $\lambda_2$ is sent along with a pump laser of wavelength $\lambda$ through a quasi-phase-matched nonlinear crystal followed by a narrowband filter. In the crystal, there is an equal probability of a photon being up- or downconverted and passing through unchanged.}
\label{fig:crystalsetup1} 
\end{figure}

Now consider the detector setup in Fig.\,\,\ref{fig:crystalsetup1}. An incoming photon with wavelength $\lambda_1$ or $\lambda_2$ as well as a pump laser beam of wavelength $\lambda$ are incident on a quasi-phase-matched nonlinear crystal.
An incoming photon of wavelength $\lambda_1$ has some probability of upconverting to a wavelength $\lambda_2$, and similarly an incoming photon of wavelength $\lambda_2$ has some probability of passing through the quasi-phase-matched nonlinear crystal unchanged.  Let us suppose that the incoming mode of the pump is in the coherent state $|\alpha, \lambda\rangle$, where
\begin{equation}
|\alpha, \lambda \rangle := e^{- \frac{|\alpha|^2}{2}} \sum_{n=0}^\infty \frac{\alpha^n}{\sqrt{n!}} \, |n, \lambda\rangle
\end{equation}
for $\alpha \in \mathbb{C}$.  In the above equation, $|n, \lambda\rangle$ is the number state in which $n$ photons of wavelength $\lambda$ are in an incoming mode.  The average number of photons in the incoming modes is given by $\langle \alpha, \lambda|\, \widehat{n} \,|\alpha, \lambda\rangle = |\alpha|^2$.  We assume that there are a large number of photons in the incoming modes, as is the case when the pump is a strong laser.

The frequency upconversion process works as follows.  Suppose that a photon of wavelength $\lambda_1$ is headed towards the crystal and that there are also incoming photons from the $\lambda$ pump.  Then the initial state of the system is $|\lambda_1\rangle |\alpha, \lambda\rangle$.
When the photon and pump enter the crystal, a nonlinear interaction occurs that causes a photon from the pump mode to upconvert the incoming photon with some probability.  This will only occur when the crystal is quasi-phase-matched for the desired process; that is to say, it has been engineered in such a way that the phase relationships are preserved allowing for the nonlinear process to be efficient.  We have
\begin{equation} \label{eq:photon1}
|\lambda_1\rangle |\alpha, \lambda\rangle \,\longrightarrow \cos(\theta) \, |\lambda_1\rangle |\alpha,\lambda\rangle +  e^{i \phi} \, \sin(\theta) \, |\lambda_2\rangle |\alpha, \lambda\rangle
\end{equation}
for some $\theta, \phi$.  Notice that in the second term, in which upconversion occurred, the $|\alpha,\lambda\rangle$ term is unchanged.  Although the coherent state has lost a photon, 
the state is unchanged since it is an eigenvector of the annihilation operator.  
Applying a $\lambda_2$ filter corresponds to the projection $|\lambda_2\rangle \langle \lambda_2 | \otimes \textbf{1}$ which leaves us with $|\lambda_2\rangle |\alpha, \lambda\rangle $. We assume that $\lambda_1$ and $\lambda_2$ are sufficiently far away from one another that the $\lambda_2$ filter completely removes light at wavelength $\lambda_1$.

If instead the incoming photon has wavelength $\lambda_2$, then we have the similar process
\begin{equation} \label{eq:photon2}
|\lambda_2\rangle |\alpha, \lambda\rangle \,\longrightarrow e^{-i \phi }\sin(\theta) \, |\lambda_1\rangle |\alpha, \lambda\rangle  +  \,\cos(\theta) \,|\lambda_2\rangle |\alpha, \lambda\rangle.
\end{equation}
For the first term of the above equation, in which downconversion occurred, the $|\alpha,\lambda\rangle$ term is unchanged even though the coherent state has gained a photon.  
This is an approximation, since the norm squared of the overlap of the modified coherent state with the original coherent state is $1 - \frac{1}{1 + \langle n \rangle}$, rather than 1.   In the limit where the number $\langle n \rangle$ of photons in the original coherent state is large, which is the case of interest for us, the approximation is accurate.  Applying the $\lambda_2$ filter to the right hand side of Eqn.  (\ref{eq:photon2}), we obtain $|\lambda_2\rangle |\alpha, \lambda\rangle$.


So far, we have seen that either an incoming photon of wavelength $\lambda_1$ or an incoming photon of wavelength $\lambda_2$ can transition to the state $|\lambda_2\rangle |\alpha, \lambda\rangle$. The final states are indistinguishable, so
if we let each detector in our HBT setup be the apparatus of Fig. \ref{fig:crystalsetup1}, we can recover interference between photons of different wavelengths.  We can use this information to infer the positions and geometries of the sources, as detailed in Section \ref{sec:resolution}.

We note from Eqn.'s~\eqref{eq:photon1} and~\eqref{eq:photon2} that for $\theta = \pi/4$, where the states $|\lambda_1\rangle$ and $|\lambda_2\rangle$ both have an equal probability of ending up in $|\lambda_2\rangle$, we can successfully project onto the $|\lambda_2\rangle$ state only half of the time.  This is optimal because we achieve the highest rate of coincidence counts when 
the transition probability between wavelengths is equal to the non-transition probability.  
It is possible to obtain additional information about the system by utilizing other values of $\theta$.  For example, $\theta = \pi/2$ corresponds to filtering out all of the light of wavelength $\lambda_1$, which gives us standard HBT with the $\lambda_2$ source alone.

A possible issue with this single crystal methodology is that one may want to choose $E_1$ and $E_2$ sufficiently close together such that the wavelength $\lambda$ corresponding to $\Delta E$ is deep in the infrared, which cannot be achieved with available nonlinear crystals.   In the next section we propose a modified procedure which remedies that problem.

\subsection{Two Crystal Method} \label{sec:twocrystal}
For situations where finding a crystal that works with a pump of wavelength $\lambda$ is not feasible, it is still possible to
upconvert photons of energies $E_1$ and $E_2$ to a common energy $E_3$ which is greater than both $E_1$ and $E_2$.  In particular, we can choose $E_3$ such that photons with energies $E_3 - E_1$ and $E_3 - E_2$ are in the optical range.

\begin{figure}[h]
\centering
\includegraphics[scale=.7]{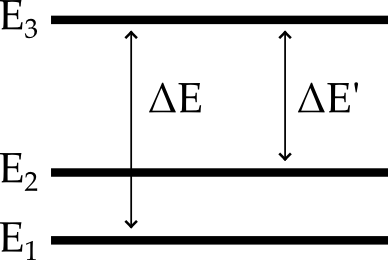} 
\caption[] {Energy gaps between photons of wavelength $\lambda_1$, $\lambda_2$, and $\lambda_3$.}
\label{fig:energygap1} 
\end{figure}

Consider photons of wavelength $\lambda_1$, $\lambda_2$, $\lambda_3$ with energies $E_1$, $E_2$, $E_3$, respectively.  Supposing that $E_1 < E_2 < E_3$, we define $\Delta E := E_3 - E_1$ and $\Delta E' := E_3 - E_2$, as shown in Fig. \ref{fig:energygap1}. 
Using similar notation as before, photons with energy $\Delta E$ will be denoted by their wavelength $\lambda$ and photons with energy $\Delta E'$ will be denoted by their wavelength $\lambda'$. \\

\newpage
Suppose that we have two quasi-phase-matched nonlinear crystals end to end, the first which upconverts $\lambda_1$ to $\lambda_3$ with a pump of wavelength $\lambda$, and the second which upconverts $\lambda_2$ to $\lambda_3$ with a pump of wavelength $\lambda'$.  Suppose that the pumps with wavelengths $\lambda$ and $\lambda'$ are phase-locked.  We can design the experiment so that both upconversion processes happen with probability very close to one.  These two processes are distinguishable since an upconverted $\lambda_1$ photon will end up in a different spatio-temporal mode than an upconverted $\lambda_2$ photon due to the the staggering of the crystals.  Let us label the initial input spatio-temporal mode by $|0\rangle$ and the two possible output spatio-temporal modes by $|1\rangle$ and $|2\rangle$.  This gives the evolution 
\begin{align}
|\lambda_1\rangle | 0 \rangle |\alpha,\lambda\rangle |\beta, \lambda'\rangle \, &\longrightarrow \, |\lambda_3\rangle | 1 \rangle |\alpha,\lambda\rangle |\beta, \lambda'\rangle \\ \nonumber \\
|\lambda_2\rangle| 0 \rangle |\alpha,\lambda\rangle |\beta, \lambda'\rangle \, &\longrightarrow \, |\lambda_3\rangle | 2 \rangle |\alpha,\lambda\rangle |\beta, \lambda'\rangle.
\end{align}

If we couple the output spatio-temporal modes to two spatial modes via an optical switch and then subsequently combine those spatial modes with a 50:50 beamsplitter, we can project onto the coherent superposition $|3\rangle = \frac{1}{\sqrt{2}} (|1\rangle + |2\rangle)$ by post-selecting on having a photon in a particular output port of the beamsplitter.  This corresponds to projecting onto $\textbf{1} \otimes |3\rangle \langle 3| \otimes \textbf{1} \otimes \textbf{1}$ where $|3\rangle$ is the state of the output port.  We are left with $|\lambda_3\rangle |3\rangle |\alpha, \lambda\rangle |\beta, \lambda'\rangle$ for both of the possible input wavelengths when the coherent states contain a large number of photons.  

Just as before, this method projects both possible input states onto the same final state and thus erases knowledge of the original wavelength. Similar schemes with two stacked crystals have previously been used to erase knowledge of other degrees of freedom during upconversion \cite{VanDevender1}.
%

\subsection{Reference Source Method}\label{sec:refmethod}

Finally we consider an alternate method, which implements E$^2$I$^2$ in a different way.  Suppose that in addition to having red and blue sources labeled $1$ and $2$, respectively, there is also a \textit{reference} source labeled $3$, which emits spatially entangled red and blue photons.  In particular, it emits a coherent superposition of states where a red photon heads towards detector $A$ while a blue photon heads towards detector $B$ \textit{and} with states where a red photon heads towards detector $B$ while a blue photon heads towards detector $A$.  One can create such a source by pumping a quasi-phase-matched nonlinear crystal that performs spontaneous parametric downconversion.  

\begin{figure}[t]
\centering
\includegraphics[scale=0.6]{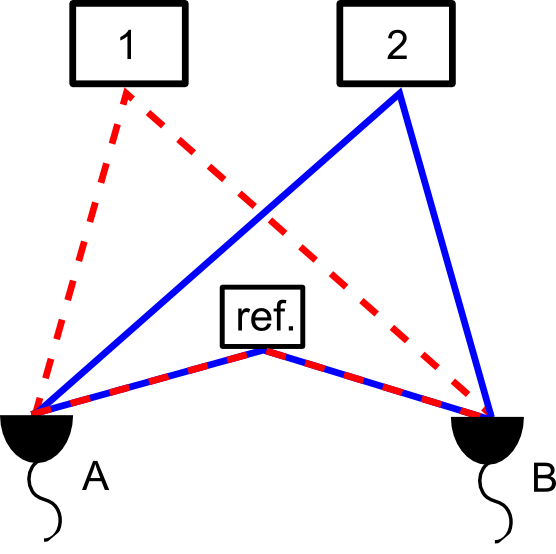} 
\caption[]{Hanbury-Brown Twiss interferometry with an additional reference source, which emits spatially entangled pairs of photons. After post-selection, the detectors cannot distinguish between receiving photons from sources 1 and the reference and sources 2 and the reference.}
\label{fig:refSource} 
\end{figure}

Consider the diagram in Fig.\,\ref{fig:refSource}.  As before, we are interested in extracting phase information about the propagation of photons from sources $1$ and $2$ to detectors $A$ and $B$ by erasing which-path information.  Sources $1$ and $2$ can be distant from the laboratory or difficult to control (such as stars or biological specimens), but the reference source must be precisely controlled by the experimenter.  Instead of making the photons from $1$ and $2$ indistinguishable by projecting them into the same state, we will instead ``confuse" the detectors by sending in additional photons from the reference source.  In particular, we will post-select on $A$ and $B$ each receiving one red photon and one blue photon, so the detectors will lose the ability to distinguish the paths traveled by the photons.
Similar to before, we define $D_{1A}$ be the propagator from source $1$ to detector $A$ and define all other propagators analogously.  By post-selecting on receiving one red photon and one blue photon at each detector, we obtain the state
\begin{equation*}
\label{newmethod1}
\left(D_{1A} D_{2B} D_{3A}^{\text{red}} D_{3B}^{\text{blue}} + D_{1B} D_{2A} D_{3B}^{\text{red}} D_{3A}^{\text{blue}} \right)\,|\redr \!  \blueb \!\,, \redr \!  \blueb \rangle.
\end{equation*}
Suppose, for simplicity, that $D_{1A} = \frac{1}{\sqrt{2}} \, e^{i \phi_{1A}}$ and similarly for the other propagators.  The $1/\sqrt{2}$ factor means that it is equally likely for a particle to go from a particular source to either detector $A$ or $B$, and the phase term is simply the accumulated phase of the photon along the designated path.  In this situation, the probability of a four-fold coincidence simplifies to
\begin{align}
\label{newmethod2}
&\left|D_{1A} D_{2B} D_{3A}^{\text{red}} D_{3B}^{\text{blue}} + D_{1B} D_{2A} D_{3B}^{\text{red}} D_{3A}^{\text{blue}}\right|^2 \nonumber \\
&\qquad \qquad \qquad \quad = \frac{1}{8} + \frac{1}{8}\,\cos\left(\phi_{1A} - \phi_{1B} - \phi_{2A} + \phi_{2B} - \phi_{3A}^{\text{red}} + \phi_{3B}^{\text{red}} + \phi_{3A}^{\text{blue}} - \phi_{3B}^{\text{blue}} \right). \nonumber \\
\end{align}
Suppose that the reference source is characterized so that we already know $\phi_{3A}^{\text{red}}$, $\phi_{3B}^{\text{red}}$, $\phi_{3A}^{\text{blue}}$, and $\phi_{3B}^{\text{blue}}$.  Then we can use Eqn. (\ref{newmethod2}) to obtain the combination of phases $\phi_{1A} - \phi_{1B} - \phi_{2A} + \phi_{2B}$, as per the standard HBT experiment.


We note that the reference source approach is very general.  In principle it works even if the two sources under consideration are photon and electron sources, respectively.  In that case, the reference source approach allows us to work around fermion-boson superselection rules (see Appendix).  Note that the reference source must generate spatially entangled photons and electrons.  That sounds more exotic than it is: indeed, it is the generic outcome of photon-electron scattering.

\section{Proposal for Implementation} \label{sec:implementation}

In this section we design, in meaningful detail, a proof-of-principle experiment implementing the first of the three methodologies detailed above. We will first describe our detector, and then the entire experimental setup containing two such detectors. Correlations in the firing of the detectors will demonstrate intensity interferometry by photons of two different wavelengths.  

Each detector contains a crystal designed such that both the up- and downconversion processes will occur with equal probability. Once the photons from the sources arrive at the detector, they are overlapped with the pump at a dichroic mirror.  The pump and source photons are then sent through the quasi-phase-matched nonlinear crystal where the frequency conversion processes occur. The pump power can be adjusted so that both processes are 50\% efficient and thus it will be impossible to tell from which source the photon came.  The crystal is placed within a cavity in order to generate a high enough pump power for the upconversion to occur. The out-coupling mirror of the cavity will be coated to transmit all of the light at $\lambda_1$ and $\lambda_2$ while reflecting the pump at $\lambda$. The output light from the cavity passes through a narrow band-pass filter, which  eliminates any remaining light at a wavelength other than $\lambda_2$.  The $\lambda_2$ photons are then detected by a common single-photon detector, such as a silicon avalanche photodiode (APD). 

We now supply some details appropriate to a demonstration experiment. For concreteness, suppose the two sources emit wavelengths of $\lambda_1$ = 780 nm and $\lambda_2$ = 519 nm, wavelengths at which diode lasers are available.  The pump would then need to have a wavelength of $\lambda$ = 1550 nm, a common telecom wavelength. For these wavelengths, there are commercially available periodically-poled lithium niobiate (PPLN) crystals which can be adapted for the desired frequency conversion processes.

\begin{figure}[t]
\centering
\includegraphics[scale=0.75]{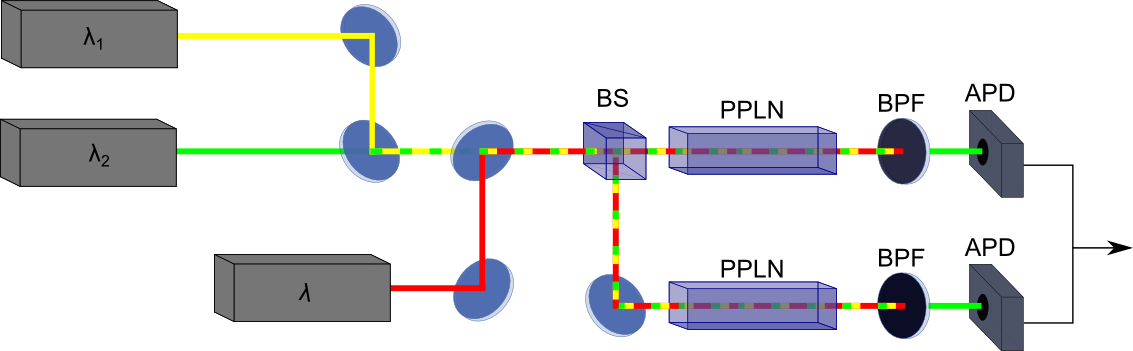} 
\vskip.3cm
\caption[]{Entire setup of this proof-of-principle experiment. Two lasers at wavelengths $\lambda_1$ (yellow) and $\lambda_2$ (green) represent the two sources for the HBT interferometry. These signal lasers are overlapped with each other and the pump of wavelength $\lambda$ (red) at dichroic mirrors. This light is split at a 50:50 beamsplitter (BS) where each part is sent to one of the detectors. Each detector consists of a PPLN cyrstal (PPLN) where the frequency conversion occurs, followed by a band-pass filter at $\lambda_2$ (BPF) and then an APD. The correlations of the counts on the two detectors are analyzed together.}
\label{fig:wholesetup} 
\end{figure}

Fig. \ref{fig:wholesetup} shows the entire experimental setup. Suppose each source is a laser with the wavelengths given above. Light from these source lasers is combined at a dichroic mirror before being overlapped with the pump at another dichroic mirror. Light at the three wavelengths are then sent together through a 50:50 beamsplitter in order to simulate the spatial separation that occurs naturally when the experimenter has no control over the sources. Each output mode is sent to one of the two detectors. Since the pump for both crystals comes from a common source, there is a consistent phase reference for the frequency conversion processes. In each detector, the source photon has a 50\% probability of exiting the crystal with a wavelength $\lambda_2$ in which case it passes through the filter and is detected. The interference term in Eqn. \ref{int1} will appear in the correlations of the intensity measured by the two detectors even though the two sources are at very different wavelengths.

Several variations on this setup are possible. For example, it would be possible to do an experiment of this kind with lower pump power, by using sources that propagate through waveguides, instead of in free space.   To demonstrate the method of Section \ref{sec:twocrystal}, one needs different detectors.    Each detector could consist of two short quasi-phase-matched nonlinear crystals back-to-back, each phase-matched for one of the two processes.  A second pump laser would be phase-locked with the first and combined with the other three wavelengths at another dichroic mirror. This setup would allow for the resulting frequency converted photons to be indistinguishable even when $E_2 - E_1$ is outside the optical spectrum.

\section{Enhanced Resolution via Intensity Correlations} \label{sec:resolution}

The original HBT experiment entailed measuring two-particle intensity correlations of the star Sirius to determine its diameter \cite{HBT1}.  More generally, people have used HBT to determine the shapes of thermal sources as well as to distinguish sources.  HBT allows one to resolve the distance between two sources with similar peak emission frequencies, even if the distance between them is less than can be resolved by each detector individually.  However, this technique fails to resolve sources with peak emission frequencies that are very different from one another.  We will show that our generalization of HBT solves this problem, thus enabling increased resolution of different-wavelength sources.

To orient the discussion, we will first provide an overview of standard HBT interferometry and then generalize the analysis to our setup with two and then many sources.

\subsection{Intensity Correlation Function for a Single Wavelength Source} \label{sec:HBTsingle}

First we will discuss the van Cittert-Zernike formula for HBT intensity correlations of an extended source emitting one wavelength of light \cite{MandelWolf1, Mansuripur1, Pearce1}.  We will assume that the source has a small coherence length.  Consider an extended source $1$ which emits light of wavelength $\lambda_{1}$ and which occupies a spatial volume $V_{1}$.  Say that we have two detectors $A$ and $B$, positioned at $\textbf{r}_A$ and $\textbf{r}_B$, respectively.  We define the phase $\Delta \phi = \Delta \phi(\textbf{r}, \lambda ; \, \textbf{r}_A, \textbf{r}_B)$ by \cite{Mansuripur1}
\begin{equation}
\Delta \phi(\textbf{r}, \lambda ; \, \textbf{r}_A, \textbf{r}_B) := \frac{2\pi}{\lambda} \left(|\textbf{r}-\textbf{r}_A| - |\textbf{r} - \textbf{r}_B| \right).
\end{equation}
Physically, $\Delta \phi(\textbf{r}, \lambda ; \, \textbf{r}_A, \textbf{r}_B)$ is the phase difference between the path of a particle of wavelength $\lambda$ propagating from $\textbf{r}$ to $\textbf{r}_A$, and a particle of wavelength $\lambda$ propagating from $\textbf{r}$ to $\textbf{r}_B$.

Let $I_{1}(\textbf{r})$ be the intensity distribution of source $1$, where $I_{1}(\textbf{r})$ vanishes outside of a finite region $V_1$ since the source has a finite spatial extent.  Defining \cite{Mansuripur1}
\begin{equation}
\widetilde{\gamma}_{1}(\textbf{x}, \textbf{y}) := \int_{V_{1}} d\textbf{r} \, I_{1}(\textbf{r}) \, e^{i \Delta \phi(\textbf{r}, \lambda_{1} ; \, \textbf{x}, \textbf{y})}
\end{equation}
we have that the two-point intensity correlation function 
is \cite{Pearce1}
\begin{equation}
\label{SingleBosonicSourceHBT}
G_{1}^{(2)}(\textbf{r}_A, \textbf{r}_B) = \widetilde{\gamma}_{1}(\textbf{r}_A, \textbf{r}_A) \, \widetilde{\gamma}_{1}(\textbf{r}_B, \textbf{r}_B) + \widetilde{\gamma}_{1}(\textbf{r}_A, \textbf{r}_B) \, \widetilde{\gamma}_{1}(\textbf{r}_B, \textbf{r}_A).
\end{equation}
In an experiment, the observed two-point intensity correlation function is equal to $G_{1}^{(2)}(\textbf{r}_A, \textbf{r}_B)$ up to an overall normalization.

To make the above formula concrete, let us compute the HBT intensity correlations for the star Sirius.  From the perspective of an observer a large distance $L$ from Sirius, the star is a disc-like source $\mathscr{D}$ with radius $a$ and angular diameter $\vartheta = 2\arctan(a/L) \approx 2a/L$.  We therefore examine the case of a disc-like source with uniform intensity.  Say that our detectors $A$ and $B$ are located at $\textbf{r}_A = (x_A, y_A, 0)$ and $\textbf{r}_B = (x_B, y_B,0)$, respectively.  We will also suppose that the star is centered at $\textbf{c} = (c_1,c_2,L)$.  If an arbitrary point on the surface of the star is $(x,y,L)$, then for large $L$ we have
\begin{align}
&\Delta \phi(x,y, \lambda ; \, x_A, y_A ; \, x_B, y_B) \approx \frac{2\pi}{\lambda L} \left(\frac{1}{2}(x_A^2 + y_A^2) - \frac{1}{2}(x_B^2 + y_B^2) - \left[(x_A - x_B) \,x + (y_A - y_B) \, y  \right] \right).
\end{align} 
In this limit,
\begin{equation}
\label{GammaEq1}
\widetilde{\gamma}_{\mathscr{D}}(\textbf{r}_A, \textbf{r}_B) = e^{-i \, \frac{2\pi}{\lambda L} (\textbf{r}_A - \textbf{r}_B)\cdot \textbf{c}} \,f_{\mathscr{D}}(\textbf{r}_A, \textbf{r}_B) 
\end{equation}
with
\begin{equation}
f_{\mathscr{D}}(\textbf{r}_A, \textbf{r}_B) := e^{i \frac{\pi}{\lambda L} (|\textbf{r}_A|^2 - |\textbf{r}_B|^2)}\, \frac{2 J_1\left(\frac{2\pi}{\lambda}\,\frac{1}{2} \vartheta \, |\textbf{r}_A - \textbf{r}_B| \right)}{\left(\frac{2\pi}{\lambda}\,\frac{1}{2} \vartheta \, |\textbf{r}_A - \textbf{r}_B| \right)}
\end{equation}
where $J_1(z)$ is the Bessel function of the first kind \cite{Pearce1}.

\begin{figure}
\centering
\includegraphics[scale=0.6]{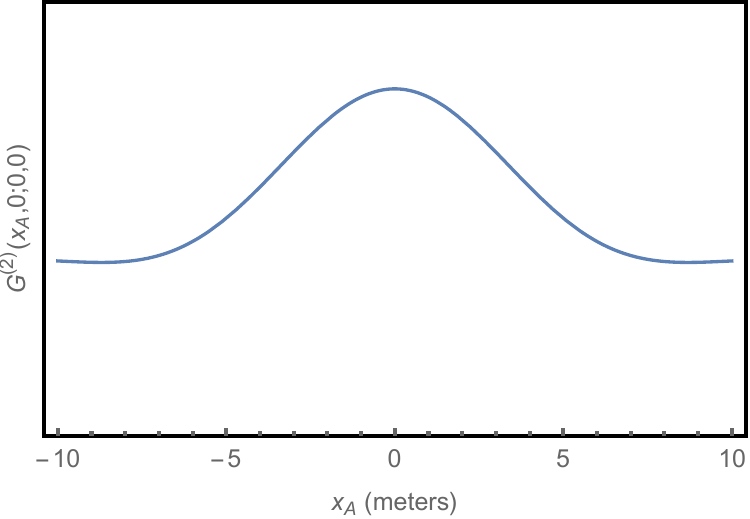} 
\caption[]{ Intensity correlation function $G_{\mathscr{D}}^{(2)}(x_A, 0; 0,0)$ for Sirius in arbitrary units as a function of the distance between the detectors.}
\label{fig:Sirius1} 
\end{figure}

Using Eqn. (\ref{SingleBosonicSourceHBT}), we can construct the two-point intensity correlation function $G_{\mathscr{D}}^{(2)}(\textbf{r}_A, \textbf{r}_B) = G_{\mathscr{D}}^{(2)}(x_A, y_A ; \, x_B, y_B)$.  It has the form
\begin{equation}
G_{\mathscr{D}}^{(2)}(\textbf{r}_A,\textbf{r}_B) = 1 + |f_{\mathscr{D}}(\textbf{r}_A, \textbf{r}_B)|^2.
\end{equation}
We have plotted $G_{\mathscr{D}}^{(2)}(x_A, 0; \, 0,0)$ in Fig. \ref{fig:Sirius1} where $L \approx 8.611$ light years is the distance from Sirius to the Earth, $a = 2 \times 10^6$ km is the diameter of Sirius, and $\lambda \approx 292$ nm is the peak blackbody wavelength of Sirius. We can determine the diameter of Sirius by measuring how quickly the peak decays as $|x_a|$ increases.

\newpage
\subsection{Intensity Correlations for Two Sources of Different Wavelength}
Now suppose that we have two sources $1$ and $2$ with different wavelengths $\lambda_1$ and $\lambda_2$.  We define
\begin{equation}
\widetilde{\gamma}_{k}(\textbf{x}, \textbf{y}) := \int_{V_{k}} d\textbf{r} \, I_{k}(\textbf{r}) \, e^{i \Delta \phi(\textbf{r}, \lambda_{k} ; \, \textbf{x}, \textbf{y})}
\end{equation}
for $k=1,2$.  Since the sources are distinguishable, their joint two-point intensity correlation function is
\begin{align}
\label{DistBosSources}
G_{\cancel{\text{E}^2 \text{I}^2}}^{(2)}(\textbf{r}_A, \textbf{r}_B) &= \left(\widetilde{\gamma}_{1}(\textbf{r}_A, \textbf{r}_A) + \widetilde{\gamma}_{2}(\textbf{r}_A, \textbf{r}_A)\right) \left(\widetilde{\gamma}_{1}(\textbf{r}_B, \textbf{r}_B) + \widetilde{\gamma}_{2}(\textbf{r}_B, \textbf{r}_B)\right) \nonumber \\
& \qquad + \widetilde{\gamma}_{1}(\textbf{r}_A, \textbf{r}_B) \, \widetilde{\gamma}_{1}(\textbf{r}_B, \textbf{r}_A) + \widetilde{\gamma}_{2}(\textbf{r}_A, \textbf{r}_B) \, \widetilde{\gamma}_{2}(\textbf{r}_B, \textbf{r}_A).
\end{align}
The $\left(\widetilde{\gamma}_{1}(\textbf{r}_A, \textbf{r}_A) + \widetilde{\gamma}_{2}(\textbf{r}_A, \textbf{r}_A)\right) \left(\widetilde{\gamma}_{1}(\textbf{r}_B, \textbf{r}_B) + \widetilde{\gamma}_{2}(\textbf{r}_B, \textbf{r}_B)\right)$ terms correspond to the correlations between double counts at individual detectors, which do not describe interference.  The terms $ \widetilde{\gamma}_{1}(\textbf{r}_A, \textbf{r}_B) \, \widetilde{\gamma}_{1}(\textbf{r}_B, \textbf{r}_A) + \widetilde{\gamma}_{2}(\textbf{r}_A, \textbf{r}_B) \, \widetilde{\gamma}_{2}(\textbf{r}_B, \textbf{r}_A)$ come from the correlations of coincidence counts between the two detectors and are interference terms because we cannot fundamentally determine from which point source each photon emerged since photons of the same wavelength are indistinguishable.  Since we can distinguish between different wavelengths of light, there are no interference terms in Eqn. (\ref{DistBosSources}) between light with different wavelengths.

\begin{figure}[t]
\centering
\subfigure[]{
\includegraphics[scale=.55]{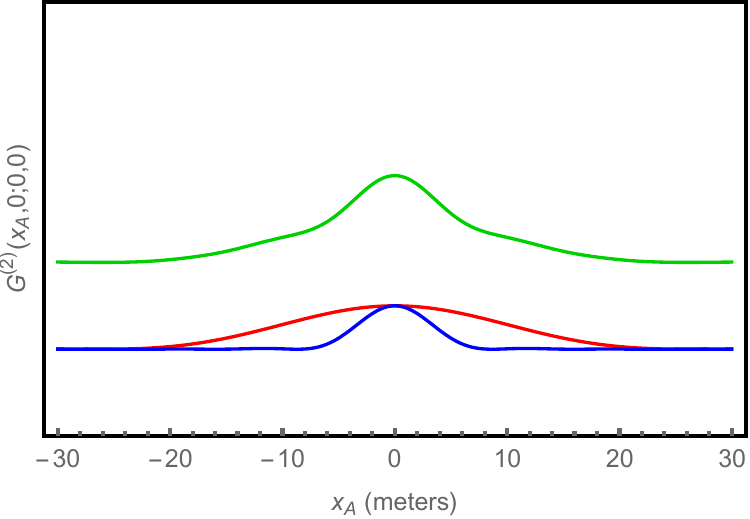}
\label{fig:noInterference}
}
\subfigure[]{
\includegraphics[scale=.55]{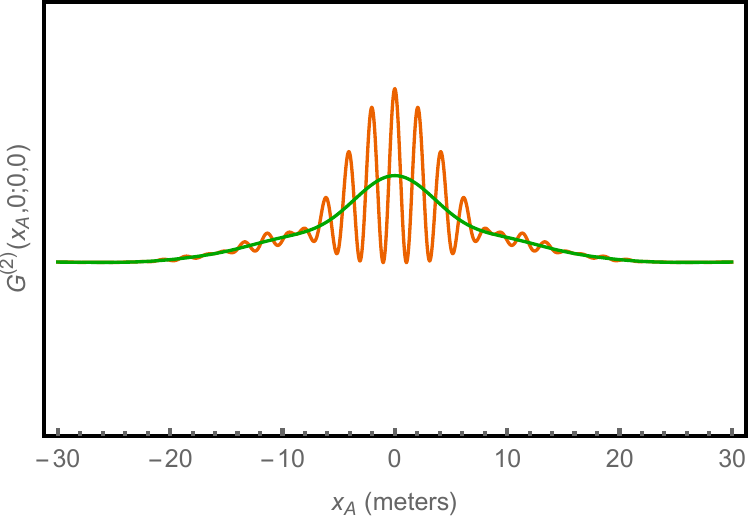}
\label{fig:interference}
}
\caption[]{Intensity correlation functions for a constructed example of two stars, both with a diameter equal to that of Sirus, both the same distance from the Earth as Sirius, and separated by a distance of four times their diameter. The first has a peak wavelength of 292 nm (the same as Sirius) and the other has a peak wavelength around 828 nm. (a) The two sources do not interfere, and the correlation function $G_{\cancel{\text{E}^2 \text{I}^2}}^{(2)}(x_A, 0; 0,0)$ is plotted in green with its contribution from each of the two sources in blue and red. (b) The intensity correlation function $G_{\text{E}^2 \text{I}^2}^{(2)}(x_A, 0; 0,0)$ is plotted in orange, compared to $G_{\cancel{\text{E}^2 \text{I}^2}}^{(2)}(x_A, 0; 0,0)$ plotted in green for which there is no interference between the two sources.} \label{fig:SiriusMultWL}
\end{figure}

If we assume the sources are uniform discs, then the intensity correlation function takes the form
\begin{equation}
G_{\cancel{\text{E}^2 \text{I}^2}}^{(2)}(\textbf{r}_A, \textbf{r}_B) = 4 + |f_1(\textbf{r}_A, \textbf{r}_B)|^2 +  |f_2(\textbf{r}_A, \textbf{r}_B)|^2.
\end{equation}
This function is plotted in Fig. \ref{fig:noInterference}.  The lack of interference between the sources prevents us from using $G_{\cancel{\text{E}^2 \text{I}^2}}^{(2)}(\textbf{r}_A, \textbf{r}_B)$ to determine the distance between the sources.  However, by filtering out all wavelengths except that of a given source, we can still determine the diameter of that source by examining how quickly its intensity correlation bump decays, as explained in Section \ref{sec:HBTsingle}.

With the aid of E$^2$I$^2$, we can recover interference between the sources of different wavelengths, thereby helping us to better resolve the distance between the two sources.  
By applying our technique of converting the wavelengths of the two sources to a common wavelength so they interfere, the joint two-point intensity correlation function
is
\begin{align}
\label{IndistBosSources}
G_{\text{E}^2 \text{I}^2}^{(2)}(\textbf{r}_A, \textbf{r}_B) &= \left(\widetilde{\gamma}_{1}(\textbf{r}_A, \textbf{r}_A) + \widetilde{\gamma}_{2}(\textbf{r}_A, \textbf{r}_A)\right) \left(\widetilde{\gamma}_{1}(\textbf{r}_B, \textbf{r}_B) + \widetilde{\gamma}_{2}(\textbf{r}_B, \textbf{r}_B)\right) \nonumber \\
& \qquad + \left(\widetilde{\gamma}_{1}(\textbf{r}_A, \textbf{r}_B) + \widetilde{\gamma}_{2}(\textbf{r}_A, \textbf{r}_B)\right) \left(\widetilde{\gamma}_{1}(\textbf{r}_B, \textbf{r}_A) + \widetilde{\gamma}_{2}(\textbf{r}_B, \textbf{r}_A)\right)
\end{align}
which notably involves interference with both wavelengths $\lambda_{1}$ and $\lambda_{2}$.  Note that Eqn. (\ref{IndistBosSources}) is just Eqn. (\ref{DistBosSources}) plus the terms $\widetilde{\gamma}_{1}(\textbf{r}_A, \textbf{r}_B) \widetilde{\gamma}_{2}(\textbf{r}_B, \textbf{r}_A)$ and $\widetilde{\gamma}_{2}(\textbf{r}_A, \textbf{r}_B) \widetilde{\gamma}_{1}(\textbf{r}_B, \textbf{r}_A)$.  Assuming that we have two uniform disc sources, we have
\begin{align}
G_{\text{E}^2 \text{I}^2}^{(2)}(\textbf{r}_A, \textbf{r}_B) &= 4 + |f_1(\textbf{r}_A, \textbf{r}_B)|^2 +  |f_2(\textbf{r}_A, \textbf{r}_B)|^2  \nonumber \\
& \qquad \qquad + 2 \,\text{Re}\left( e^{-i \,\frac{2\pi}{L}\, (\textbf{r}_A - \textbf{r}_B) \cdot \left(\frac{1}{\lambda_1} \, \textbf{c}_1 - \frac{1}{\lambda_2} \, \textbf{c}_2 \right)} f_1(\textbf{r}_A, \textbf{r}_B) f_2^*(\textbf{r}_A, \textbf{r}_B)\right).
\end{align}
This intensity correlation function is plotted in Fig. \ref{fig:interference}. 

The oscillations in Fig.\,\,\ref{fig:interference} are a signature of two distinct sources, and the frequency of these oscillations can be used to determine the distance between them.  If the distance between the sources is $d$, then the oscillations are due to terms of the form
$$\exp\left(\pm i \pi\,\frac{d}{L}\, |\textbf{r}_A - \textbf{r}_B| \left(\frac{1}{\lambda_1} + \frac{1}{\lambda_2}\right)\right)$$
and so the frequency of the oscillations goes as $\frac{d}{L} \cdot \frac{1}{2}\left(\frac{1}{\lambda_1} + \frac{1}{\lambda_2} \right)$.  Therefore, E$^2$I$^2$ allows us to utilize the increased resolving power of two-point intensity correlation functions over one-point intensity functions to better resolve the distance between two sources of different wavelength.
It is worth emphasizing that our methodology works even when the sources of interest are mutually incoherent, since phases intrinsic to the sources cancel out in the HBT interference terms.

\begin{figure}[t]
\centering
\includegraphics[scale=0.6]{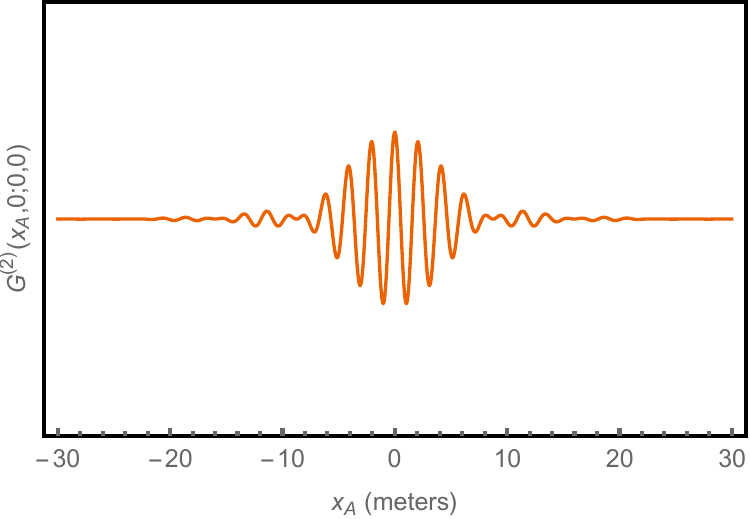} 
\caption[]{ Intensity correlation function $G_{\text{E}^2\text{I}^2}^{(2)}(x_A, 0; 0,0)$ of the same setup as in Figure \ref{fig:SiriusMultWL}.}
\label{fig:Sirius2} 
\end{figure}

A natural quantity to consider is the two-point intensity correlation function in the presence of E$^2$I$^2$ minus the two-point intensity correlation function in the absence of E$^2$I$^2$.  Thus, we define
\begin{align}
\widetilde{G}_{\text{E}^2 \text{I}^2}^{(2)}(\textbf{r}_A, \textbf{r}_B) :&= G_{\text{E}^2 \text{I}^2}^{(2)}(\textbf{r}_A, \textbf{r}_B) - G_{\cancel{\text{E}^2 \text{I}^2}}^{(2)}(\textbf{r}_A, \textbf{r}_B) \\
&= \widetilde{\gamma}_{1}(\textbf{r}_A, \textbf{r}_B) \, \widetilde{\gamma}_{2}(\textbf{r}_B, \textbf{r}_A) + \widetilde{\gamma}_{2}(\textbf{r}_A, \textbf{r}_B) \, \widetilde{\gamma}_{1}(\textbf{r}_B, \textbf{r}_A).
\end{align}
For the disc sources previously mentioned, we have
\begin{equation}
\widetilde{G}_{\text{E}^2 \text{I}^2}^{(2)}(\textbf{r}_A, \textbf{r}_B) = 2 \,\text{Re}\left( e^{-i \,\frac{2\pi}{L}\, (\textbf{r}_A - \textbf{r}_B) \cdot \left(\frac{1}{\lambda_1} \, \textbf{c}_1 - \frac{1}{\lambda_2} \, \textbf{c}_2 \right)} f_1(\textbf{r}_A, \textbf{r}_B) f_2^*(\textbf{r}_A, \textbf{r}_B)\right).
\end{equation}
This function is plotted in Fig. \ref{fig:Sirius2}.
It is clear from the above equations that $\widetilde{G}_{\text{E}^2 \text{I}^2}^{(2)}(\textbf{r}_A, \textbf{r}_B)$ captures the non-trivial interference induced by E$^2$I$^2$.  By determining $\widetilde{G}_{\text{E}^2 \text{I}^2}^{(2)}(\textbf{r}_A, \textbf{r}_B)$ via measurements, one can take its Fourier transform to extract $\frac{1}{\lambda_1} \, \textbf{c}_1 - \frac{1}{\lambda_2} \, \textbf{c}_2$ which yields information above the relative positions of the sources.

\subsection{Generalization to Many Sources}
Suppose we now have $n$ sources, where the sources labeled $1,...,k$ have wavelengths \newline $\lambda_1,...,\lambda_k = \lambda$ and the sources labeled $k+1,...,n$ have wavelengths $\lambda_{k+1},...,\lambda_n = \lambda'$.  Then we have the two-point intensity correlation functions
\begin{align}
G_{\cancel{\text{E}^2 \text{I}^2}}^{(2)}(\textbf{r}_A, \textbf{r}_B) &= \left(\sum_{i=1}^n \widetilde{\gamma}_i(\textbf{r}_A, \textbf{r}_A) \right)\left( \sum_{i=1}^n \widetilde{\gamma}_i(\textbf{r}_B, \textbf{r}_B) \right) + \left( \sum_{i=1}^k \widetilde{\gamma}_i(\textbf{r}_A, \textbf{r}_B)  \right)\left( \sum_{i=1}^k \widetilde{\gamma}_i(\textbf{r}_B, \textbf{r}_A)  \right) \nonumber \\
&\qquad \qquad \qquad \qquad \qquad \qquad \qquad \qquad \,\,\,+ \left( \sum_{i=k+1}^n \widetilde{\gamma}_i(\textbf{r}_A, \textbf{r}_B)  \right)\left( \sum_{i=k+1}^n \widetilde{\gamma}_i(\textbf{r}_B, \textbf{r}_A)  \right) \\
\label{TwoPointFunction1}
G_{\text{E}^2 \text{I}^2}^{(2)}(\textbf{r}_A, \textbf{r}_B) &= \left(\sum_{i=1}^n \widetilde{\gamma}_i(\textbf{r}_A, \textbf{r}_A) \right)\left( \sum_{i=1}^n \widetilde{\gamma}_i(\textbf{r}_B, \textbf{r}_B) \right) + \left( \sum_{i=1}^n \widetilde{\gamma}_i(\textbf{r}_A, \textbf{r}_B)  \right)\left( \sum_{i=1}^n \widetilde{\gamma}_i(\textbf{r}_B, \textbf{r}_A)  \right)
\end{align}
\begin{align}
\widetilde{G}_{\text{E}^2 \text{I}^2}^{(2)}(\textbf{r}_A, \textbf{r}_B) &= \left( \sum_{i=1}^k \widetilde{\gamma}_i(\textbf{r}_A, \textbf{r}_B)  \right)\left( \sum_{i=k+1}^n \widetilde{\gamma}_i(\textbf{r}_B, \textbf{r}_A)  \right) + \left( \sum_{i=k+1}^n \widetilde{\gamma}_i(\textbf{r}_A, \textbf{r}_B)  \right)\left( \sum_{i=1}^k \widetilde{\gamma}_i(\textbf{r}_B, \textbf{r}_A)  \right).
\end{align}
Notice that $\widetilde{G}_{\text{E}^2 \text{I}^2}^{(2)}(\textbf{r}_A, \textbf{r}_B)$ is only comprised of interference terms between the sources of wavelength $\lambda$ and the sources of wavelength $\lambda'$.  For the case of uniform discs, we find the particularly interesting equation
\begin{align}
G_{\text{E}^2 \text{I}^2}^{(2)}(\textbf{r}_A, \textbf{r}_B) = n^2 + \sum_{i=1}^n |f_i(\textbf{r}_A, \textbf{r}_B)|^2 + \sum_{p <  q} 2 \,\text{Re}\left( e^{-i \,\frac{2\pi}{L}\, (\textbf{r}_A - \textbf{r}_B) \cdot \left(\frac{1}{\lambda_p} \, \textbf{c}_p - \frac{1}{\lambda_q} \, \textbf{c}_q \right)} f_p(\textbf{r}_A, \textbf{r}_B) f_q^*(\textbf{r}_A, \textbf{r}_B)\right).
\end{align}
By taking its Fourier transform with respect to $\textbf{r}_A$ and $\textbf{r}_B$, we can recover information about the relative positions between each pair of sources by knowing $\frac{1}{\lambda_p} \, \textbf{c}_p - \frac{1}{\lambda_q} \, \textbf{c}_q$ for all pairs of sources $p,q$.

\section{Applications} \label{sec:applications}

Having seen how E$^2$I$^2$ can be used to resolve nearby sources of different spectral character, here we discuss several areas where this capability could be helpful.  There are many possible applications of E$^2$I$^2$ for different wavelengths, involving a wide range of scales.  

In general, prior to use of E$^2$I$^2$ one should first measure a standard 1-point intensity function to locate the sources of interest approximately (for instance, by imaging).  If the 1-point function is insufficient to resolve the sources, one can consider performing an E$^2$I$^2$ 2-point function measurement, with the detectors aimed at the sources of interest.

Fluorescence microscopy is a technique pervasive in the biological sciences \cite{FM1, FM2}.  In this technique, organic specimens are labeled with phosphorescent dyes which emit at different frequencies.  Typically different dyes are preferentially absorbed by functionally distinct biological units (e.g. macromolecules or organelles).   The biological sample is then illuminated, and one images the organic specimens using light emitted by the dyes.  When two dyes of different wavelengths are absorbed by nearby units, E$^2$I$^2$ can provide enhanced resolving power.   
Since some versions of fluorescence microscopy project onto a single plane, one unit may occlude others.  By tilting the plane of the detectors one can effectively turn the problem of resolving one dye behind the other into a problem of resolving two dyes that are close together, which is a problem E$^2$I$^2$ can solve \cite{Kaldun}.   More detailed exploration of fluorescence microscopy as well as the imaging of crystals is currently in progress \cite{Kaldun}.

Another application is to the imaging of astrophysical sources.  Since E$^2$I$^2$ methodology could augment HBT, and it piggybacks naturally on HBT telescopes, it should be considered in conjunction with new proposed HBT experiments \cite{HBTastro1, HBTastro2, HBTastro3}.

\section{Conclusion}

We have presented a practical method to implement E$^2$I$^2$ with sources of different wavelengths of light, and showed how this can be used to better resolve sources of different spectral character.  Since the HBT effect has proved fruitful in a wide variety of applications, we anticipate useful applications of our multiple-wavelength generalization.

These developments exemplify a broader strategy, whereby one sacrifices the ability to extract some information about a quantum system by projecting otherwise orthogonal outcomes onto a common final state, enabling interference.  In the spirit of complementarity, different measurement protocols bring out different, mutually exclusive aspects of the system under consideration.

\subsection*{Acknowledgements}
The authors would like to thank Philip Bucksbaum, Savas Dimopolous, Martin Fejer, Aram Harrow, Yu-Ping Huang, Andreas Kaldun, Mark Kasevich, Tim Kovachy, Ken Van Tilburg and Vladan Vuleti\'{c} for valuable conversations and feedback.  JC is supported by the Fannie and John Hertz Foundation and the Stanford Graduate Fellowship program. FW is supported by the Swedish Research Council under Contract No. 335-2014-7424, the European Research Council under grant 742104, and the U.S. Department of Energy under grant Contract  No. DE-SC0012567.  VB is supported by the National Science Foundation Graduate Research Fellowship under grant number DGE-114747.  

\newpage

\section*{Appendix}

\subsection*{A \quad Polarization}

Above, we have focused on sources which emit photons of different wavelengths.  Even if the sources emit photons of the same wavelength, the question naturally arises: If the sources have non-trivial polarization properties, can we access them? 
For example: If we have two very nearby sources, that emit in orthogonal polarizations, can we resolve them?  
Unadorned HBT will not serve here, but (as we shall see) a simple refinement accesses much more information, and does the job.  

Let us first consider the simple case where source 1 produces photons with polarization described, in a basis of orthogonal linear polarizations, by 
$\left(\begin{array}{c}\alpha \\ \beta \end{array}\right)$, while source 2 produces photons with polarization $\left(\begin{array}{c}\gamma \\ \delta\end{array}\right)$.  Furthermore, let us apply projections $\Pi_A, \Pi_B$ at the two detectors.   Then the rate for simultaneous firing becomes, as in our earlier discussion, a sum of the separate process terms
\begin{eqnarray}
&{}&\left(\begin{array}{cc}\alpha^* & \beta^*\end{array}\right) \Pi_A   \left(\begin{array}{c}\alpha \\\beta\end{array}\right)  \ \,
\left(\begin{array}{cc}\gamma^* & \delta^*\end{array}\right)  \Pi_B  \left(\begin{array}{c}\gamma \\\delta\end{array}\right)\  |D_{1A}|^2 |D_{2B}|^2 \ +\nonumber \\
&{}&
\left(\begin{array}{cc}\gamma^* & \delta^*\end{array}\right)  \Pi_A  \left(\begin{array}{c}\gamma \\\delta\end{array}\right)  \ \, 
\left(\begin{array}{cc}\alpha^* & \beta^*\end{array}\right)  \Pi_B  \left(\begin{array}{c}\alpha \\\beta\end{array}\right) \ |D_{2A}|^2 |D_{1B}|^2 
\end{eqnarray}
and the interference term
\begin{equation}
\left(\begin{array}{cc}\gamma^* & \delta^*\end{array}\right)  \Pi_A  \left(\begin{array}{c}\alpha \\\beta\end{array}\right)  \ \, 
\left(\begin{array}{cc}\alpha^* & \beta^*\end{array}\right)  \Pi_B  \left(\begin{array}{c}\gamma \\\delta\end{array}\right)\ \, 
D_{1A}D_{2B}D_{2A}^*D_{1B}^* \ + \ {\rm c.c.}
\end{equation}

We can generalize this by allowing the sources to emit in mixtures, described by polarization (density) matrices $\pi_1, \pi_2$.  Then we get for the uncrossed terms
\begin{equation}\label{direct}
{\rm Tr} \,  \Pi_A  \pi_1  \, {\rm Tr} \,  \Pi_B  \pi_2 \ |D_{1A}|^2 |D_{2B}|^2 \, + \,
{\rm Tr} \,  \Pi_A  \pi_2  \, {\rm Tr} \,  \Pi_B  \pi_1 \  |D_{2A}|^2 |D_{1B}|^2 
\end{equation}
and for the crossed term
\begin{eqnarray}\label{crossed}
&{}& {\rm Tr} \, \Pi_A \pi_1 \Pi_B \pi_2 \ D_{1A}D_{2B}D_{2A}^*D_{1B}^* \, + \, {\rm c.c.}  \\
&=& {\rm Tr} \, \Pi_A \pi_1 \Pi_B \pi_2 \ D_{1A}D_{2B}D_{2A}^*D_{1B}^* \, + \, {\rm Tr} \, \Pi_A \pi_2 \Pi_B \pi_1 \ D_{1A}^*D_{2B}^*D_{2A}D_{1B} \nonumber
\end{eqnarray}
where in the second line we exploit the hermiticity of $\pi, \Pi$.   

By letting the $\Pi$s interpolate between the two orthogonal polarizations of the sources in our model problem, we obtain interference between them, which could allow us to resolve them.   More generally, use of the $\Pi$s can significantly enhance our perception of the sources.   

With $\Pi_A = \Pi_B = \textbf{1}$ intensity interferometry accesses the {\it cross-polarization}, an interesting quantity that has been discussed previously \cite{Wolf1, Wolf2, Wolf3}.  We call the more general phenomenon ``linked polarization.''   

Note that in this procedure we have {\it gained \/} one form of information by {\it erasing\/} potential information that would have enabled us, in principle, to say which source was responsible for the emission (even when we cannot resolve it spatially).   That erasure renders two otherwise distinguishable processes to become indistinguishable, and enables their interference.  This is the essence of E$^2$I$^2$.

\subsection*{B \quad Bosons and Fermions}
\subsubsection*{B.1 \quad Introducing Entanglement}

So far, we have considered recovering interference between photons of different wavelength, and polarization.  To demonstrate the scope and generality of E$^2$I$^2$, consider the extreme example that one of our sources emits bosons, while another source emits fermions. 
A detector that receives a boson goes into state {\bf B}, while a detector that receives a fermion goes into state {\bf F}.  

We would like to get interference between the terms in 
\begin{equation*}
S_{1A}\, D_{2B} | \, {\bf FB} \rangle \, + \, D_{2A} \, S_{1B} \, | {\bf BF} \rangle\,,
\end{equation*}
where $S$ denotes a fermion propagator and $D$ denotes a boson propagator.  Following the E$^2$I$^2$ philosophy, we change the state basis and erase information to access interference.

We can do that directly, using entangled detector states (Procedure 1).  Writing 
\begin{align*}
S_{1A}\, D_{2B} | \, {\bf FB} \rangle \, + \, D_{2A} \, S_{1B} \, | {\bf BF} \rangle \nonumber &= \frac{1}{2} ( S_{1A}\, D_{2B} \, + \, D_{2A} \, S_{1B} ) (| \, {\bf FB} \rangle \, + \, | {\bf BF} \rangle) \nonumber \\
& \quad + \frac{1}{2} ( S_{1A}\, D_{2B} \, - \, D_{2A} \, S_{1B} ) (| \, {\bf FB} \rangle \, - \, | {\bf BF} \rangle) \nonumber 
\end{align*}
we see that by projecting on the entangled state
\begin{equation*}
\frac{1}{\sqrt 2} ( |\,{\bf FB} \rangle \, + \, | {\bf BF} \rangle )
\end{equation*}
we measure
\begin{equation*}
| S_{1A} D_{2B} \, + \, D_{2A} S_{1A} |^2
\end{equation*}

We might also follow the polarization strategy more literally, acting on the detectors separately (Procedure 2).  Here, with 
\begin{eqnarray}
| {\bf F} \rangle_A ~&=&~ \frac{1}{\sqrt 2} ( | C \rangle \, + \, | D \rangle) \nonumber \\
| {\bf B} \rangle_A ~&=&~ \frac{1}{\sqrt 2} ( | C \rangle \, - \, | D \rangle) \nonumber \\
| {\bf F} \rangle_B ~&=&~ \frac{1}{\sqrt 2} ( | E \rangle \, + \, | F \rangle) \nonumber \\
| {\bf B} \rangle_B ~&=&~ \frac{1}{\sqrt 2} ( | E \rangle \, - \, | F \rangle) 
\end{eqnarray}
projection on 
\begin{equation*}
| C \rangle \, \langle C | \, \otimes | E \rangle \, \langle E |
\end{equation*}
gives us what we want.  

However, Procedure 2 requires that we set up coherent superpositions of states that differ by one in fermion number.   This is problematic, because it violates a superselection rule.   Procedure 1 is free of that issue, because the two parts of $\frac{1}{\sqrt 2} ( |\, {\bf FB} \rangle \, + \, | {\bf BF} \rangle )$, while they differ in fermion number locally, agree in that respect globally.   Below, we shall indicate a geometrical method for realizing this entanglement.   The possibility of measurable boson-fermion interference sheds an interesting light on superselection, emphasizing its global nature.  

\subsubsection*{B.2 \quad Entanglement by Spatial Superposition}

In Procedure 1, to measure interference between the states of a boson and fermion, we needed to project detector states onto $\frac{1}{
\sqrt{2}}(|\,\textbf{FB}\rangle + |\textbf{BF}\rangle)$.  Here we describe a more geometric way to obtain the desired interference.

Say that we have two detectors located at $A$ and $B$ respectively.  The detector located at $A$ begins in a fermion accepting state $|\mathbb{F}\rangle$ which transitions to $|\textbf{F}\rangle$ if and only if it absorbs a fermion.  Similarly, the detector at $B$ begins in a boson accepting state $|\mathbb{B}\rangle$ which transitions to $|\textbf{B}\rangle$ if and only if it absorbs a boson.  The initial state of the detectors is $|\mathbb{F}\rangle|A\rangle \otimes |\mathbb{B}\rangle|B\rangle$ where we have included states that keep track of the positions of the detectors.

Now we put the detectors in an equal superposition of being at their original positions and being in swapped positions as
\begin{equation} \label{detectorSwap}
\frac{1}{\sqrt{2}} \left(|\mathbb{F}\rangle|A\rangle \otimes |\mathbb{B}\rangle|B\rangle + |\mathbb{F}\rangle|B\rangle \otimes |\mathbb{B}\rangle|A\rangle \right)
\end{equation}
This superposition of swapped and unswapped detectors can be obtained by a unitary operation, for example application of the unitary operator
\begin{align}
V = &\frac{1}{\sqrt{2}} \left( |A\rangle \otimes |B\rangle +  |B\rangle \otimes |A\rangle \right) \left(\langle A| \otimes  \langle B| \right) + \frac{1}{\sqrt{2}} \left( |A\rangle \otimes |B\rangle -  |B\rangle \otimes |A\rangle \right) \left(\langle B| \otimes  \langle A| \right) \nonumber \\
&+ |A\rangle\langle A| \otimes |A\rangle\langle A| + |B\rangle\langle B| \otimes |B\rangle\langle B| \nonumber
\end{align}
to the spatial states.  Note that $V$ has the property $V^2 = \textbf{1}$.

If we have a fermion emitter and a boson emitter as before, we will be interested in the terms
\begin{equation} \label{detectorSwapRegistered}
S_{1A} D_{2B} |\textbf{F}\rangle |A\rangle \otimes |\textbf{B}\rangle |B\rangle + D_{2A} S_{1B}  |\textbf{F}\rangle |B\rangle \otimes |\textbf{B}\rangle |A\rangle
\end{equation}
Applying $V$ to the terms in Eqn. (\ref{detectorSwapRegistered}) we obtain
\begin{equation}
\frac{1}{\sqrt{2}} ( S_{1A} D_{2B} + D_{2A} S_{1B} ) |\textbf{F}\rangle |A\rangle \otimes |\textbf{B}\rangle |B\rangle + \frac{1}{\sqrt{2}} ( S_{1A} D_{2B} - D_{2A} S_{1B} ) |\textbf{F}\rangle |B\rangle \otimes |\textbf{B}\rangle |A\rangle\nonumber
\end{equation}
Projecting on the separable state $ |\textbf{F}\rangle |A\rangle \otimes |\textbf{B}\rangle |B\rangle$, we measure $|S_{1A}D_{2B} + D_{2A}S_{1B}|^2$ as desired.

In practice, of course, maintaining quantum coherence through a spatial swap operation presents a demanding challenge, but at least it is a clearly defined one.

\subsection*{C \quad General Entanglement}

We can generalize E$^2$I$^2$ for HBT in the following way.   The projectors $\Pi_A, \Pi_B$ implement measurement by the detectors $A, B$.  When we want to make an entangled measurement, however, we should not utilize a tensor product of projectors $\Pi_A \otimes \Pi_B$ but instead a joint projector ${\bf \Pi}_{AB}$.  Written in index notation, we are replacing
\begin{equation*}
(\Pi_A)^{\alpha_1}_{\alpha_2} \, (\Pi_B)^{\beta_1}_{\beta_2} \, \rightarrow \, {\bf \Pi}^{\alpha_1 \beta_1}_{\alpha_2 \beta_2}
\end{equation*}
where we have suppressed the $AB$ subscript of ${\bf \Pi}_{AB}$ to avoid notational clutter.

We should also allow for the interesting possibility of entanglement of the sources.  The density matrices describing the sources need not factorize, and so we can replace 
\begin{equation*}
(\pi_1)^{\alpha_1}_{\alpha_2} \, (\pi_2)^{\beta_1}_{\beta_2} \, \rightarrow \, {\boldsymbol \pi}^{\alpha_1 \beta_1}_{\alpha_2 \beta_2}
\end{equation*}

With these notations, we can generalize our formula Eqn.\,(\ref{direct}) in the form
\begin{equation}\label{genDirect}
{\bf \Pi}^{\alpha_1 \beta_1}_{\alpha_2 \beta_2} \, {\boldsymbol \pi}^{\alpha_2 \beta_2}_{\alpha_1 \beta_1} \, |D_{1A}|^2 |D_{2B}|^2 \, + \,
 {\bf \Pi}^{\alpha_1 \beta_1}_{\alpha_2 \beta_2} \, {\boldsymbol \pi}^{\beta_2 \alpha_2}_{\beta_1 \alpha_1} \,  |D_{2A}|^2 |D_{1B}|^2 
 \end{equation}
and our formula Eqn.\,(\ref{crossed}) in the form
\begin{equation}\label{genCrossed}
{\bf \Pi}^{\alpha_1\beta_1}_{\alpha_2 \beta_2} \, {\boldsymbol \pi}_{\alpha_1 \beta_1}^{\beta_2 \alpha_2}  \ D_{1A}D_{2B}D_{2A}^*D_{1B}^* \, 
+ \, 
{\bf \Pi}^{\alpha_1\beta_1}_{\alpha_2 \beta_2} \, {\boldsymbol \pi}_{\beta_1 \alpha_1}^{\alpha_2 \beta_2} \ D_{1A}^*D_{2B}^*D_{2A}D_{1B}
\end{equation}

By comparing experimental data with Eqn.'s\,(\ref{genDirect}) and  (\ref{genCrossed}) we become sensitive to entanglement between the emitters.   If we know that the source is a pure state, we need not check whether
${\boldsymbol \pi}$ factorizes.  Otherwise, we need to check whether ${\boldsymbol \pi}$ can be expressed as a sum of tensor products of positive semi-definite Hermitian matrices (i.e., candidate density matrices), which is more complex algebraically.  This criterion could be used as a probe for proposed exotic states of matter that feature long-range entanglement, e.g. by studying their fluorescence.   It would also be interesting to investigate the possible existence of entanglement and linked polarization in the microwave background radiation.  

\subsection*{D \quad Single Source of Decaying Particles}

A central theme of E$^2$I$^2$ is measuring detector states in non-standard bases.  We can leverage this technique in situations when there is only a single source.  Here we will discuss the interesting case of a single source that is a collection of identical decaying particles.

Consider a particle $X$ which either decays into two particles of type $Y$ or two particles of type $Z$.  Thus, the two decay channels are $X \to YY$ and $X \to ZZ$ which occur with probability amplitudes $\mathcal{M}_{X \to YY}$ and $\mathcal{M}_{X \to ZZ}$ respectively.  Typically, when we compute information about a decay process we are interested in the absolute squares of the probability amplitudes $|\mathcal{M}_{X \to YY}|^2$, $|\mathcal{M}_{X \to ZZ}|^2$ which are measured in standard particle experiments.  By considering mixed bases of detector states, we gain access to the relative phases between the probability amplitudes.

Let ``$1$" denote a source of decaying $X$ particles.   Additionally, we have two detectors at $A$ and $B$ which are in $YY$ and $ZZ$ accepting states respectively.  We want interference between the terms
\begin{equation}
\mathcal{M}_{X \to YY} D_{1A} |\textbf{YY}\rangle |\mathbb{ZZ}\rangle + \mathcal{M}_{X \to ZZ} D_{1B} |\mathbb{YY}\rangle |\textbf{ZZ}\rangle\,.  \nonumber
\end{equation}
Projecting onto $\frac{1}{\sqrt{2}} (|\textbf{YY}\rangle |\mathbb{ZZ}\rangle + |\mathbb{YY}\rangle |\textbf{ZZ}\rangle) $ we obtain
\begin{equation}
|\mathcal{M}_{X \to YY} D_{1A} + \mathcal{M}_{X \to ZZ} D_{1B}|^2 \nonumber
\end{equation}
which gives interference between the probability amplitudes for decay.  A similar procedure can be used to measure relative phases between the probability amplitudes of scattering processes.

\end{document}